\documentstyle[amssymb,aps,12pt]{revtex}

\begin{document}
\author{M. S. Hussein, C. P. Malta, M. P. Pato and A.P.B. Tufaile}
\address{Center for Experimental and Theoretical Studies in Quantum Chaos and Related%
\\
Areas (ETC$^{*}$)\\
Instituto de F\'{\i}sica, Universidade de S\~{a}o Paulo\\
C.P. 66318, 05315-970 S\~{a}o Paulo, S.P., Brazil}
\title{Effect of symmetry breaking on level curvature distributions\thanks{%
Supported in part by the CNPq - Brazil and FAPESP}}
\maketitle

\begin{abstract}
We derive an exact general formalism that expresses the eigenvector and the
eigenvalue dynamics as a set of coupled equations of motion in terms of the
matrix elements dynamics. Combined with an appropriate model Hamiltonian,
these equations are used to investigate the effect of the presence of a
discrete symmetry in the level curvature distribution. An explanation of the
unexpected behavior of the data regarding frequencies of acoustic vibrations
of quartz block is provided.
\end{abstract}

\section{Introduction}

The usefulness of the study of statistical properties of eigenvalues and
eigenvectors of quantum systems has already been demonstrated in many areas
of physics. A lot can be learned, especially about symmetries, by just
employing the appropriate statistics. It has also become clear that these
statistics follow universal patterns that can be modelled by probability
distributions extracted from an ensemble of random Hamiltonians of the same
class of the underlying symmetry of the system under study\cite{Mehta}. This
has been a field of intense investigation over the last two decades\cite
{Guhr}. These activities have concentrated their effort on what we can call
the ``statics'' of the problem, in which stationary Hamiltonians are
considered. More recently, however, the interest has also been directed to
the dynamical aspects of the same question.

The ``dynamics'' consist in considering a given Hamiltonian as a function of
a parameter (representing ``time''). The statistical properties that
characterize the evolution are then studied as the parameter is varied. Only
evolutions that preserve the symmetry class of the Hamiltonian are
considered. Several measures have been introduced to investigate this kind
of evolution. One of the most used ones is the probability distribution of
the level curvature, which can be thought of as ``acceleration'' as it is
defined in terms of the second derivative with respect to the parameter.
These distributions measure correlations among the set of eigenvalues.
Another measure that is commonly used is the two-point correlation function
between first derivatives (``velocities''). Given some generic level, this
two-point correlation function is obtained by calculating the ``velocity''
at two different values of the parameter\cite{Simons}. Measures have also
been considered to characterize the evolution of the eigenvectors\cite
{Alhassid}.

These studies started with Wilkinson's pioneering work that investigated the
dependence of the eigenvalues of a fully chaotic billiard as a function of
its shape\cite{Wilk}. The plot of the trajectories of levels as a function
of the parameter that controls the shape, exhibits a typical pattern of
avoided crossings. A measure of these is provided by the curvature of the
trajectory. There is now an analytical evidence that, in the fully chaotic
regime, the curvatures, after an appropriate rescaling, follow an universal
simple distribution. The tail of this distribution has been investigated,
and an asymptotic dependence inversely proportional to the third power of
the curvature was established, for fully chaotic systems that are time
reversal invariant and thus governed by the Gaussian Orthogonal Ensemble
(GOE)\cite{Delande}. The expression

\begin{equation}
P\left( k\right) =\frac{1}{2\left( 1+k^{2}\right) ^{\frac{3}{2}}}  \label{1}
\end{equation}
was then proposed for the entire domain of the curvature $k$. Finally, it
has been proved that this function gives the exact distribution of
curvatures, in the case of random matrix ensembles\cite{Oppen}. The power $%
\frac{3}{2}$ in the denominator is the GOE value of $\frac{\left( \beta
+2\right) }{2}$ with $\beta =1,2$ and $4$ for the GOE, Gaussian Unitary
Ensemble (GUE) and Gaussian Symplectic Ensemble (GSE), respectively.

Recently, the difficult task of checking experimentally this prediction was
undertaken by the experimental group at the Center for Chaos and Turbulence
of the Niels Bohr Institute\cite{Clive}. They studied the dependence on the
temperature, the external parameter in the system, of the spectrum of
frequencies of quartz blocks. In previous investigations\cite{Clive0}, they
have found that the spectra of frequencies of quartz blocks obey statistical
models based on random matrix theories. The dynamics of the frequencies, as
a function of the temperature, was therefore measured for a quartz block
whose static statistical properties were already previously established. The
data obtained, however, have shown a deviation from the above expected
curvature distribution. This deviation, although slight, is significant and
not yet completely understood. We are going to show, in the present paper,
that the data can in fact be fully understood if one requires the average
curvature to be equal to one, as is implied by the universal distribution,
Eq. (\ref{1}).

So far, all studies of parametric correlations have been concentrated on the
fully chaotic regime when the system statistics are well described by the
Gaussian ensembles of Random Matrix Theory (RMT), in particular, the
Orthogonal Ensemble (GOE), if there is time-reversal invariance. The
partially chaotic situation has been little investigated. We intend here to
provide the first systematic discussion of this situation. In section II, we
develop the formalism and the model we employ, and in section III, we
present the numerical results and discussion. We verify that, at the GOE
limit, the above universal expression for the level curvature distribution
is obtained. As some degree of symmetry is introduced, it is found that the
distribution becomes narrower. However, as the symmetry is progressively
introduced, the distributions return to the universal function, if the
average of curvatures is imposed to be one.

\section{The formalism and the model}

We shall now derive a set of equations that describe simultaneously the
dynamics of the energy levels and of the eigenvector components of a
Hamiltonian $H$. These equations contain the equations of motion of the
matrix elements of $H,$ whose dependence on the parameter, $t$, representing
the ``time'', is supposed to be given. Our starting point is the general
matrix equation

\begin{equation}
H=UH_{D}U^{\dagger },  \label{2}
\end{equation}
where $H$ is an $N\times N$ real symmetric matrix, $H_{D}$ is the diagonal
matrix constructed with the $N$ eigenvalues, and $U$ is the unitary matrix
whose columns are the $N$ eigenvectors. Assuming that $H_{D}$ and $U$ also
depend on the parameter $t$, differentiating Eq. (\ref{1}) with respect to $%
t $ we get 
\begin{equation}
\stackrel{.}{H}=U\stackrel{.}{H_{D}}U^{\dagger }+\stackrel{.}{U}%
H_{D}U^{\dagger }+UH_{D}\stackrel{.}{U^{\dagger }},  \label{4}
\end{equation}
where the derivative is indicated by a dot. Multiplying $\left( \ref{4}%
\right) $ by $U^{\dagger }$ from the left, by $U$ from the right and
defining the matrix $S=U^{\dagger }\stackrel{.}{U}=-\stackrel{.}{\stackrel{.%
}{U^{\dagger }}U\text{ }}$we obtain the equation of motion

\begin{equation}
\stackrel{.}{H}_{D}=\left[ H_{D},S\right] +P  \label{6}
\end{equation}
where the matrix $P=U^{\dagger }\stackrel{.}{H}U$ was introduced. On the
other hand, we find for $P$ the conjugate evolution equation

\begin{equation}
\dot{P}=\left[ P,S\right] +U^{\dagger }\ddot{H}U  \label{7}
\end{equation}

By choosing a particular model, i.e., the dependence of the matrix element
on the parameter $t$, these equations can be employed in several contexts.
They can be used, for example, to construct an alternative method of matrix
diagonalization, or, by requiring the matrix elements to satisfy appropriate
Langevin equations, they lead to Dyson's Brownian motion model\cite{Dyson}.
Here, we concentrate on the simple model given by

\begin{equation}
H=H_{1}\cos t+H_{2}\sin t\text{ ,}  \label{8}
\end{equation}
where $H_{1}$ and $H_{2}$ are a couple of fixed, i.e., parameter
independent, random matrices taken from the same matrix ensemble, and $t$ is
the parameter. If in (\ref{8}) $H_{1}$ and $H_{2}$ are taken from a Gaussian
ensemble, the evolution will preserve the probability distribution, so that $%
H$ will remain in the same ensemble. With this choice, (\ref{7}) becomes

\begin{equation}
\dot{P}=\left[ P,S\right] -H_{D}  \label{8a}
\end{equation}

The pair of coupled equations, (\ref{6}) and (\ref{8a}), have the explicit
solution

\[
H_{D}\left( t\right) =U^{\dagger }\left( t\right) \left[ H_{D}\left(
0\right) \cos t+P\left( 0\right) \sin t\right] U\left( t\right) 
\]
and

\[
P\left( t\right) =U^{\dagger }\left( t\right) \left[ -H_{D}\left( 0\right)
\sin t+P\left( 0\right) \cos t\right] U\left( t\right) 
\]
where $U\left( t\right) $ is the solution of the equation $\stackrel{.}{U}%
=US $ given by

\[
U\left( t\right) =T\exp \int_{0}^{t}S\left( \tau \right) d\tau 
\]
with $T$ being the time-ordering operator.

To implement this solution numerically a basis has to be chosen to express
the eigencomponents. Since our objective is to investigate the curvature, a
quantity related to the behavior of the eigenvalues as the external
parameter is varied, it is convenient to use the instantaneous Hamiltonian
eigenstates as basis vectors. In this case, from the diagonal part of (\ref
{6}) we derive

\begin{equation}
\dot{E_{k}}=P_{kk}\text{,}  \label{9}
\end{equation}
and using in (\ref{7}) the relation $S_{kl}=P_{kl}/\left( E_{k}-E_{l}\right) 
$ (obtained from the off-diagonal part of (\ref{6})) we can then derive the
equations

\begin{equation}
\stackrel{.}{P_{kl}}=-\frac{P_{kk}-P_{kl}}{E_{k}-E_{l}}P_{kl}+\sum_{m=1,m%
\neq k,l}^{N}P_{km}P_{lm}\left( \frac{1}{E_{k}-E_{m}}+\frac{1}{E_{l}-E_{m}}%
\right) ,  \label{10}
\end{equation}
and

\begin{equation}
\stackrel{.}{P_{kk}}=-E_{k}+\sum_{m=1,m\neq k}^{N}\frac{2P_{km}^{2}}{%
E_{k}-E_{m}}.  \label{11}
\end{equation}

This set of coupled equations is one of the main results of this paper. All
calculations will be based on it. Thus the ``accelerations'', i.e. the
levels' curvature, are just given by (\ref{11}).

Regarding the random matrix ensemble, we shall work with a Gaussian ensemble
that interpolates between one GOE and two decoupled GOE's. This ensemble has
been already employed with a very satisfactory result in the analysis of
data relative to symmetry breaking\cite{Pato1,Guhr1} in nuclear\cite{Mitchel}
and acoustic systems\cite{Clive0}. It can be defined by the following
operator equation\cite{Pato2}

\begin{equation}
H=PH^{GOE}P+QH^{GOE}Q+\lambda \left( PH^{GOE}Q+QH^{GOE}P\right) \text{,}
\label{20}
\end{equation}
where $P=\sum\limits_{i=1}^{M}P_{i},$ $Q=1-P$ and $P_{i}=\mid i><i\mid ,$ $%
i=1,\ldots ,N$ are projection operators, $0\leq \lambda \leq 1$ is the
parameter that controls the transition, and $H^{GOE}$ denotes a GOE matrix
whose elements follow a joint probability distribution given by

\begin{equation}
P\left( H^{GOE}\right) \propto \exp \left[ -\alpha 
\mathop{\rm tr}%
\left( H^{GOE}\right) ^{2}\right] ,  \label{25}
\end{equation}
with $\alpha $ being an arbitrary scaling parameter. With the above
definitions, $\lambda =1$ corresponds to the GOE case, while $\lambda =0$
corresponds to block diagonal random matrices, made up of two GOE matrices
of sizes $M\times M$ and $\left( N-M\right) \times \left( N-M\right) $.

Regarding this ensemble, it is important to stress that $\lambda $ is not
the more convenient parameter to work with, since the transition is also
dependent on the matrix size, $N$. Independence on the dimension is obtained
by introducing the scaled parameter

\begin{equation}
\varepsilon =\sqrt{N}\lambda .  \label{28}
\end{equation}

\section{Numerical results and discussion}

Before presenting the results, we discuss the rescaling variables necessary
to extract a universal behavior. First, we have to unfold the spectrum, that
is, we work with a new spectrum generated by the transformation

\begin{equation}
x_{l}=\int_{-\infty }^{E_{l}}dE\overline{\rho }\left( E\right) \text{ for }%
l=1,...,N\text{ ,}  \label{30}
\end{equation}
where $\overline{\rho }\left( E\right) $ is the averaged level density.
Without loss of generality, we consider in the calculation only the
symmetric situation, $N=2M$, in which the matrices are decomposed into
blocks of equal size. In this case, the average density is given by the
Wigner's semicircle law\cite{Gervois}. With an appropriate scaling that
guarantees the correct value of the second moment of the eigenvalue, the
level density is given by 
\begin{equation}
\overline{\rho }\left( E\right) =\frac{4\alpha }{\pi \left( 1+\lambda
^{2}\right) }\sqrt{\frac{N}{2\alpha }\left( 1+\lambda ^{2}\right) -E^{2}}.
\label{32}
\end{equation}
In the Fig. 1, we show the nice fit obtained with this expression when
compared with the numerical values of $\overline{\rho }\left( E\right) $
generated within the two coupled GOE's ensemble alluded to above.

Second, we need some normalization of the accelerations. This is a
controversial issue that requires some discussion. On one hand, it has been
proposed that the parameter $t$ should be replaced by a new dimensionless
parameter $\tau $ related to $t$ by\cite{Leboeuf}

\begin{equation}
\frac{d\tau }{dt}=\sqrt{\left\langle \dot{x}^{2}\right\rangle }\text{ ,}
\label{36}
\end{equation}
where the average of the velocity is made over the whole set of eigenvalues
or, equivalently, over the ensemble. The level curvature is then defined in
terms of these new scaled variables as

\begin{equation}
K=\frac{1}{\pi }\frac{d^{2}x}{d\tau ^{2}}\text{ }=\frac{1}{\pi \left\langle 
\dot{x}^{2}\right\rangle }\left( \stackrel{..}{x}-\frac{\left\langle 
\stackrel{.}{x}\stackrel{..}{x}\right\rangle }{\left\langle \dot{x}%
^{2}\right\rangle }\stackrel{.}{x}\right) \text{,}  \label{40}
\end{equation}
where $\stackrel{.}{x}=\overline{\rho }\left( E\right) \dot{E}$ and $%
\stackrel{..}{x}=\overline{\rho }\left( E\right) \ddot{E}+\frac{d\overline{%
\rho }\left( E\right) }{dE}\left( \dot{E}\right) ^{2}.$

On the other hand, the universal curvature distribution, Eq. (\ref{1}),
implies that $\left\langle \left| k\right| \right\rangle =1.$ It is not at
all clear that the scaled curvatures given by equation (\ref{40}) will
satisfy this condition. Thus, we imposed the normalization

\begin{equation}
k=\frac{K}{<\left| K\right| >}  \label{41}
\end{equation}
with $K$ given by (\ref{40}). Our calculations have shown that this last
step is necessary in order to get stable results, i. e., independent of the
subset of levels of the spectra over which the statistics is performed.

The behavior of the distribution, Eq. (\ref{1}), for large curvatures can be
traced to the level spacing distribution. In fact, large curvatures can be
considered, approximately, as inversely proportional to the small level
spacing $s$. Thus if we assume $s\varpropto 1/k$ and use the fact that, in
the GOE case, $P\left( s\right) $ is linear in $s$, we obtain

\begin{equation}
P\left( k\right) \sim P\left( s\right) \left| \frac{ds}{dk}\right| \sim
k^{-3}\text{ ,}  \label{42}
\end{equation}
as predicted by $\left( \ref{1}\right) $. As a consequence, as symmetry is
introduced by decreasing the parameter $\lambda ,$ one would expect a
reduction on the probability of large curvatures with the distribution
becoming narrower. We shall see that this indeed happens.

Our main results are presented in Fig. 2, where the curvature distributions
were calculated for four values of the scaled parameter (\ref{28}). The
figures show that, at the two extreme situations, namely, in the one GOE
limit, and in the two fully decoupled GOE's limit, the curvatures distribute
themselves according to the universal distribution. We would expect this
kind of behavior in the latter limit, since the levels in each block become
completely independent of the levels in the other. Thus, their trajectories
can cross freely. As the discussion in the preceding paragraph predicted,
the distributions are narrower in the intermediate region. We stress that
these results are strongly dependent on the renormalization (\ref{41}).

Turning now to the question of the behavior (distributions wider than the
universal) presented by the data of Ref. \cite{Clive}, one possible
explanation would be that the curvatures do not average to one. To check
this point we have fitted the data with the distribution

\begin{equation}
P\left( K\right) =\frac{1}{2\gamma \left[ 1+\left( K/\gamma \right)
^{2}\right] ^{\frac{3}{2}}}  \label{46}
\end{equation}
in which the average curvature, $\gamma =<\left| K\right| >$, is treated as
a free parameter. The best fit, obtained with $\gamma =1.27\pm 0.01,$ is
displayed in the figure 3. This excellent fit makes this explanation
plausible.

In conclusion, we have investigated the effect of the symmetry breaking on
the level curvature distribution using a random matrix ensemble that allows
for the transition from one GOE to two decoupled GOE's. We have also
provided an explanation for the discrepancy\cite{Clive} of the data
regarding the temperature dependence of frequencies of acoustic vibrations
of quartz blocks.

\newpage

{\bf Figure Captions} 
\begin{figure}[tbp]
\caption{Densitiy of levels: comparison of the calculated density
(histogram) with the semi-circle law (\ref{32})\ (solid line). The
calculation corresponds to matrices of dimension $N=100,$ and $\varepsilon
=0.32.$}
\end{figure}

\begin{figure}[tbp]
\caption{Level curvature distributions: comparison of the calculated
histograms with the theoretical prediction (\ref{1})\ (solid line). The
calculations correspond to matrices of dimension $N=100,$ and for the values
of $\varepsilon$ indicated in the figure.}
\end{figure}

\begin{figure}[tbp]
\caption{Fitting of the data (crosses) of Ref. \protect\cite{Clive} with the
parametrized distribution (\ref{46})\ (solid line). The best fit (${\chi}%
^2=0.00004$) was obtained with $\gamma=1.27\pm 0.01$. The dotted curve
corresponds to the universal distribution.}
\end{figure}

\newpage \newpage

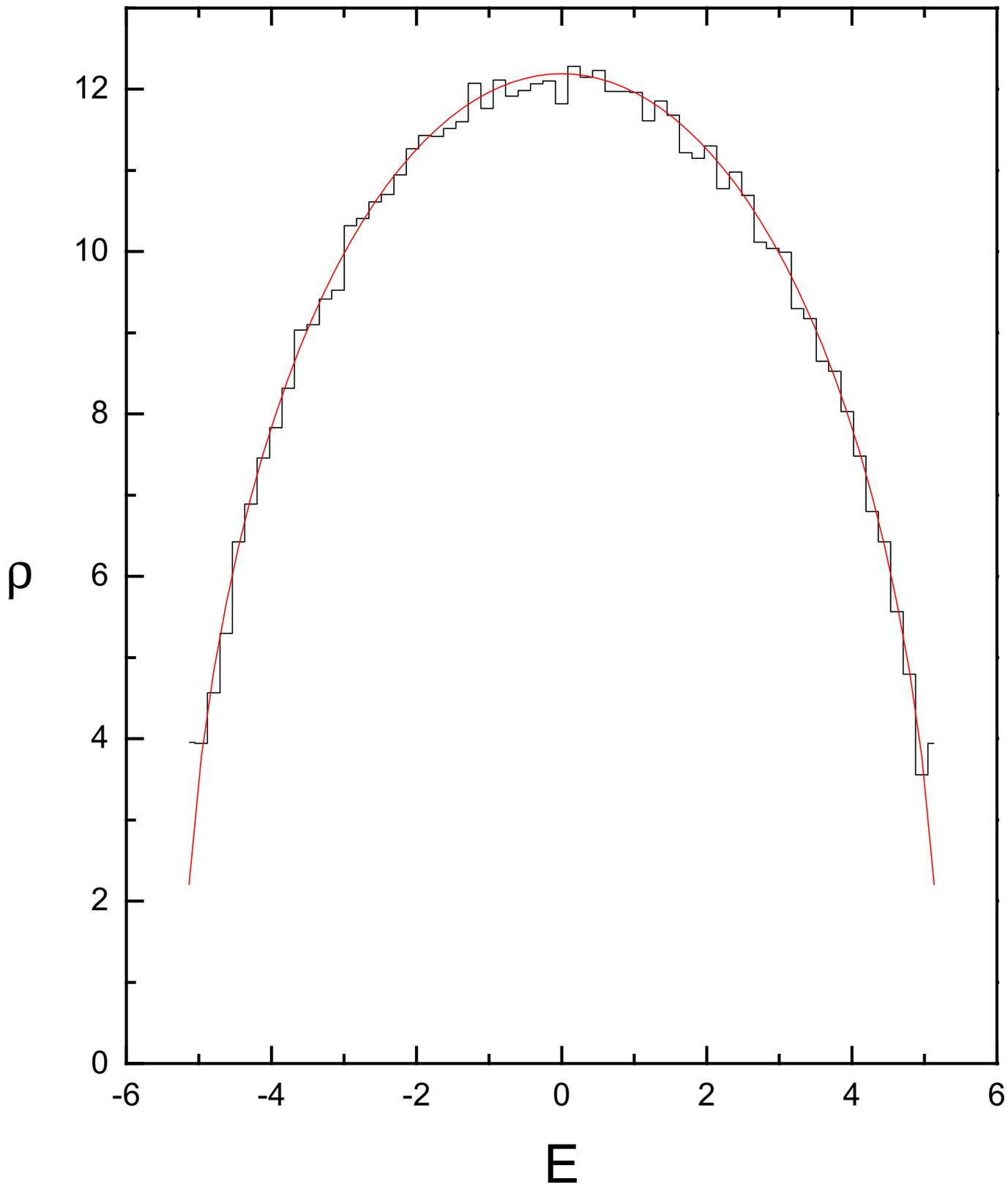
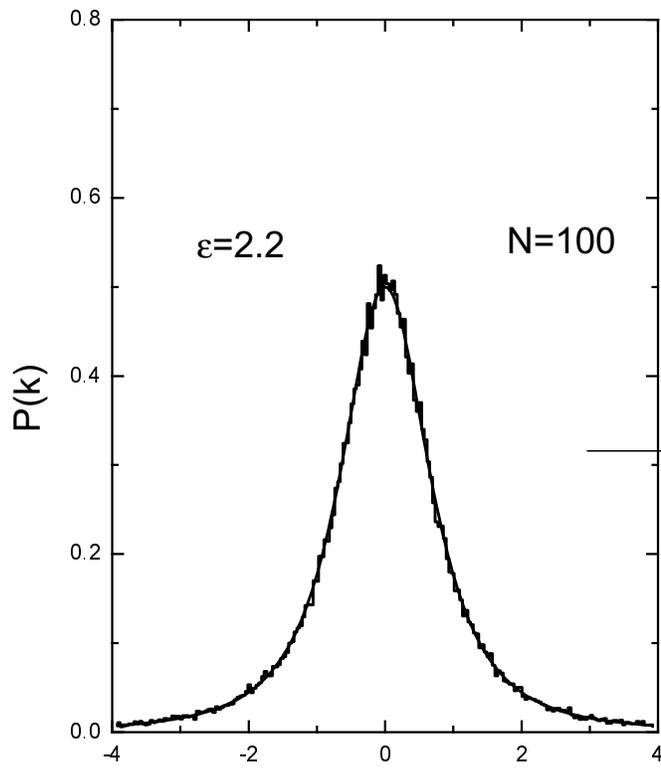
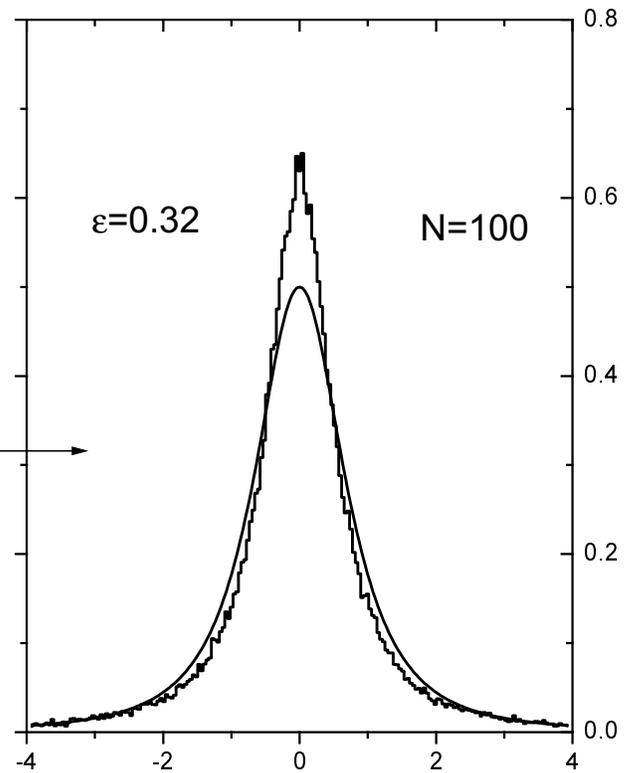
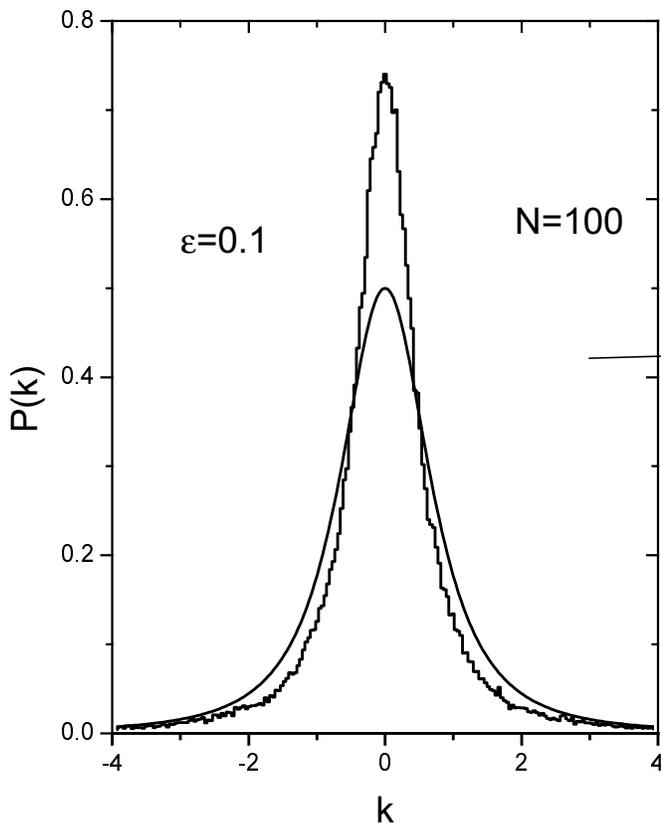
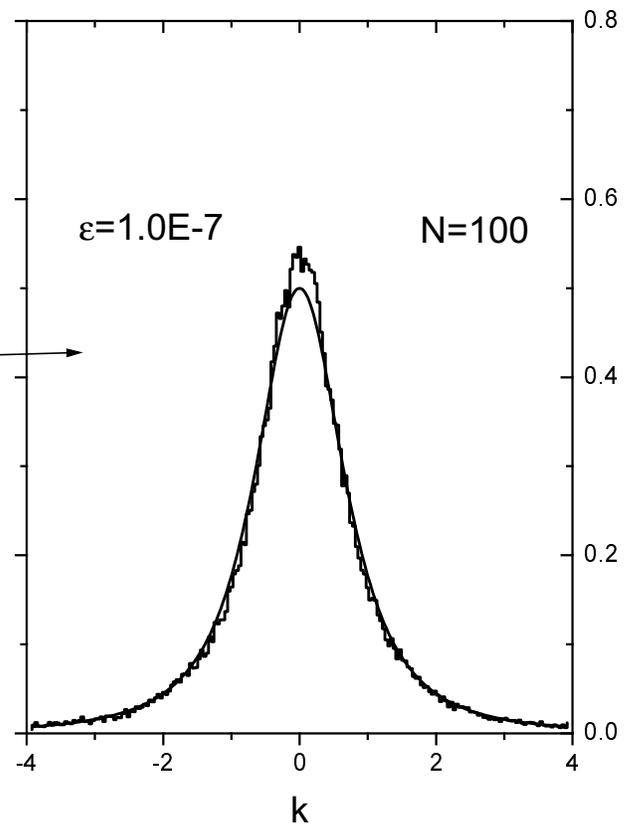
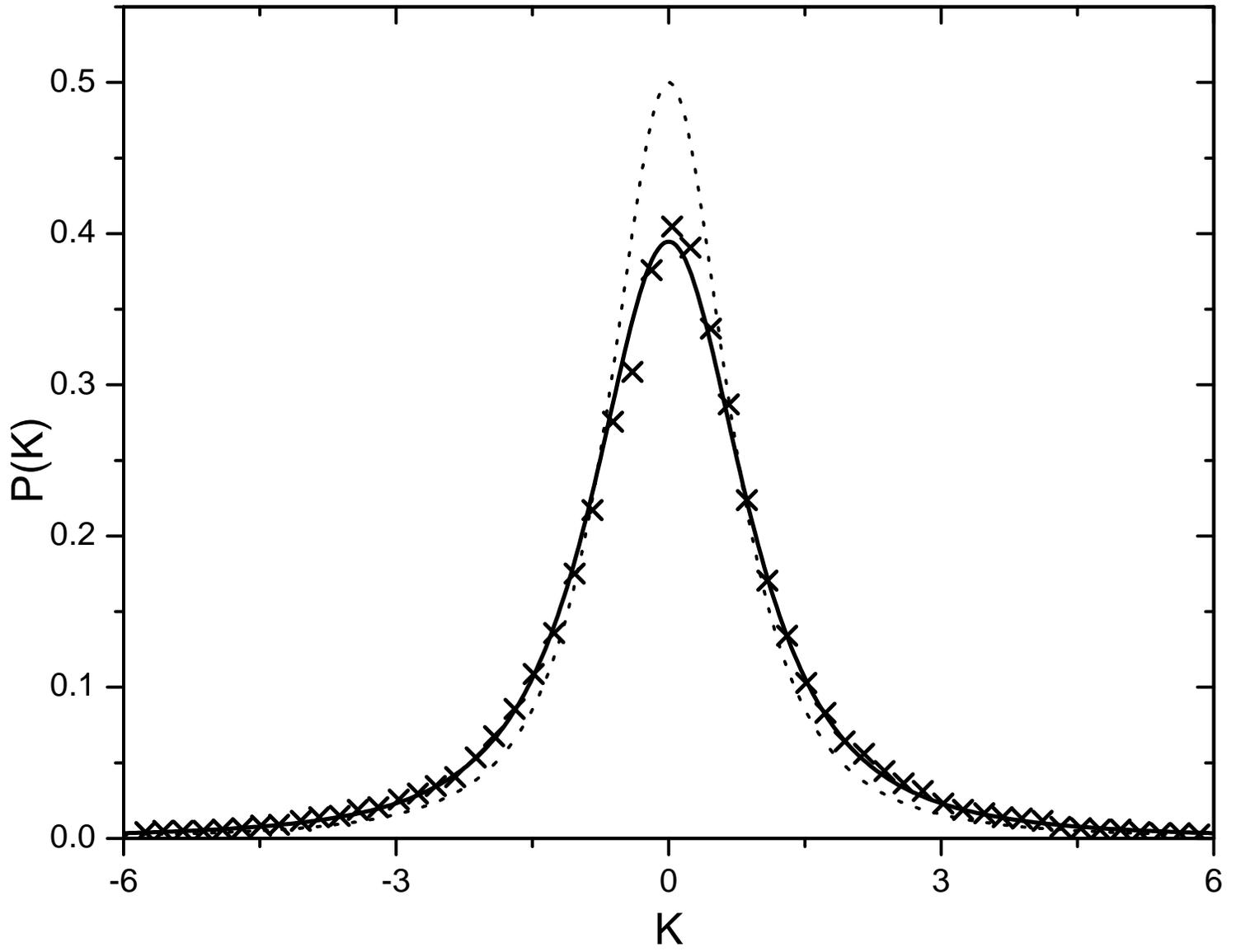
\end{document}